# First observation of Cherenkov rings with a large area CsI-TGEM-based RICH prototype

V. Peskov[a,b,*], G. Bencze[a,c], A. Di Mauro[a], P. Martinengo[a], D. Mayani[b], L. Molnar[a,c], E. Nappi[d], G. Paic[b], N. Smirnov[e], H. Anand[f], I. Shukla[g]

[a.] CERN, Geneva, Switzerland
[b.] Instituto de Ciencias Nucleares, Universidad Nacional Autonoma de Mexico, Mexico, Mexico
[c.] MTA KFKI RMKI, Research Institute for Particle and Nuclear Physics, Budapest, Hungary
[d.] Universita degli Studi di Bari, Dipartimento Interateneo di Fisica "M. Merlin" & INFN Sezione di Bari, Bari, Italy
[e.] Yale University, New Haven, USA
[f.] University of Delhi, India
[g.] Mechanical Engineering Department, NIT Durgapur, India

[*]Corresponding author:
Vladimir.peskov@cern.ch
Tel:+41227674643
Fax: +41 22 7679480

**Abstract**

We have built a RICH detector prototype consisting of a liquid $C_6F_{14}$ radiator and six triple Thick Gaseous Electron Multipliers (TGEMs), each of them having an active area of 10x10 cm$^2$. One triple TGEM has been placed behind the liquid radiator in order to detect the beam particles, whereas the other five have been positioned around the central one at a distance to collect the Cherenkov photons. The upstream electrode of each of the TGEM stacks has been coated with a 0.4 μm thick CsI layer.
    In this paper, we will present the results from a series of laboratory tests with this prototype carried out using UV light, 6 keV photons from $^{55}$Fe and electrons from $^{90}$Sr as well as recent results of tests with a beam of charged pions where for the first time Cherenkov Ring images have been successfully recorded with TGEM photodetectors. The achieved results prove the feasibility of building a large area Cherenkov detector consisting of a matrix of TGEMs.

**Introduction**

In the framework of the ALICE upgrade program we are investigating the possibility of building a new RICH detector allowing to extend the particle identification for hadrons up to 30 GeV/c. The suggested detector will consist of a gaseous radiator (for example $CF_4$ or $C_4F_{10}$) and a planar gaseous photodetector. One of the options for the photodetector which is currently under evaluation is the use of CsI coated TGEMs. The TGEM [1-3] is a hole-type gaseous multiplier based on standard printed circuit boards featuring a combination of mechanical drilling (by a CNC drilling machine) and etching techniques.

The main advantages of this approach are: TGEMs can operate at higher gas gains than CsI coated MWPC, in a large variety of gases including nonflammable ones [4] and in the same gases as those used for the Cherenkov radiator, offering the attractive possibility to build windowless RICH detectors. Note that CsI coated GEMs are currently used successfully in the PHENIX hadron blind detector [5]. TGEMs also have several attractive features compared to ordinary GEMs: a) robustness (capability to withstand sparks without being destroyed) and b) a self- supporting mechanical structure making their use convenient in large detectors. These TGEM advantages have motivated the present work.

Our first RICH prototype in which the TGEM photodetector was used was a triple CsI-TGEM combined with a $CaF_2$ radiator [6]. Beam tests of this detector demonstrated its excellent performance (such as stability and high achievable gas gains) [6]. However, since most of the Cherenkov light was trapped inside the $CaF_2$ crystal (due to internal light reflection inside the crystal) it was possible to extract the Cherenkov light detection efficiency only by simulations.

In this paper, we present the first results obtained with a large area proximity focusing TGEM RICH prototype with a $C_6F_{14}$ liquid radiator (15 mm thickness) in which there is no light trapping effect.

**Detector design**

The detector design is shown in Fig. 1. The detector consists of a 15 mm thick liquid $C_6F_{14}$ radiator and six triple TGEMs each of them having an active area of 10x10 cm$^2$ The characteristics of each TGEM are: thickness 0.45 mm, hole diameter 0.4 mm, rims thickness 10 μm and 0.8 mm pitch. A drift electrode (wire grid) was positioned 3 cm above the top electrode of the triple TGEMs. One triple TGEM was placed behind the liquid radiator in order to detect the beam particles, whereas the other five have been positioned along a circle in order to collect the Cherenkov photons. The upstream electrode of each TGEM stack has been coated with a 0.4 μm thick CsI layer. The gas chamber has windows in front of each triple TGEM allowing to irradiate the detectors either with radioactive sources such as $^{55}$Fe or $^{90}$Sr or with UV light from a Hg lamp. A pad read-out plane (pad size 8x8 mm$^2$) was placed 3 mm behind the triple TGEMs combined with a front-end electronics based on the standard ALICE-HMPID GASSIPLEX and DILOGIC chips [7]. Event monitoring and analysis were carried out using the ALICE software packages AliRoot and AMORE [8].

There was also the possibility to independently observe analog signals from any of the electrodes of the TGEM stacks and, if necessary, to optimize voltages on any particular TGEM.

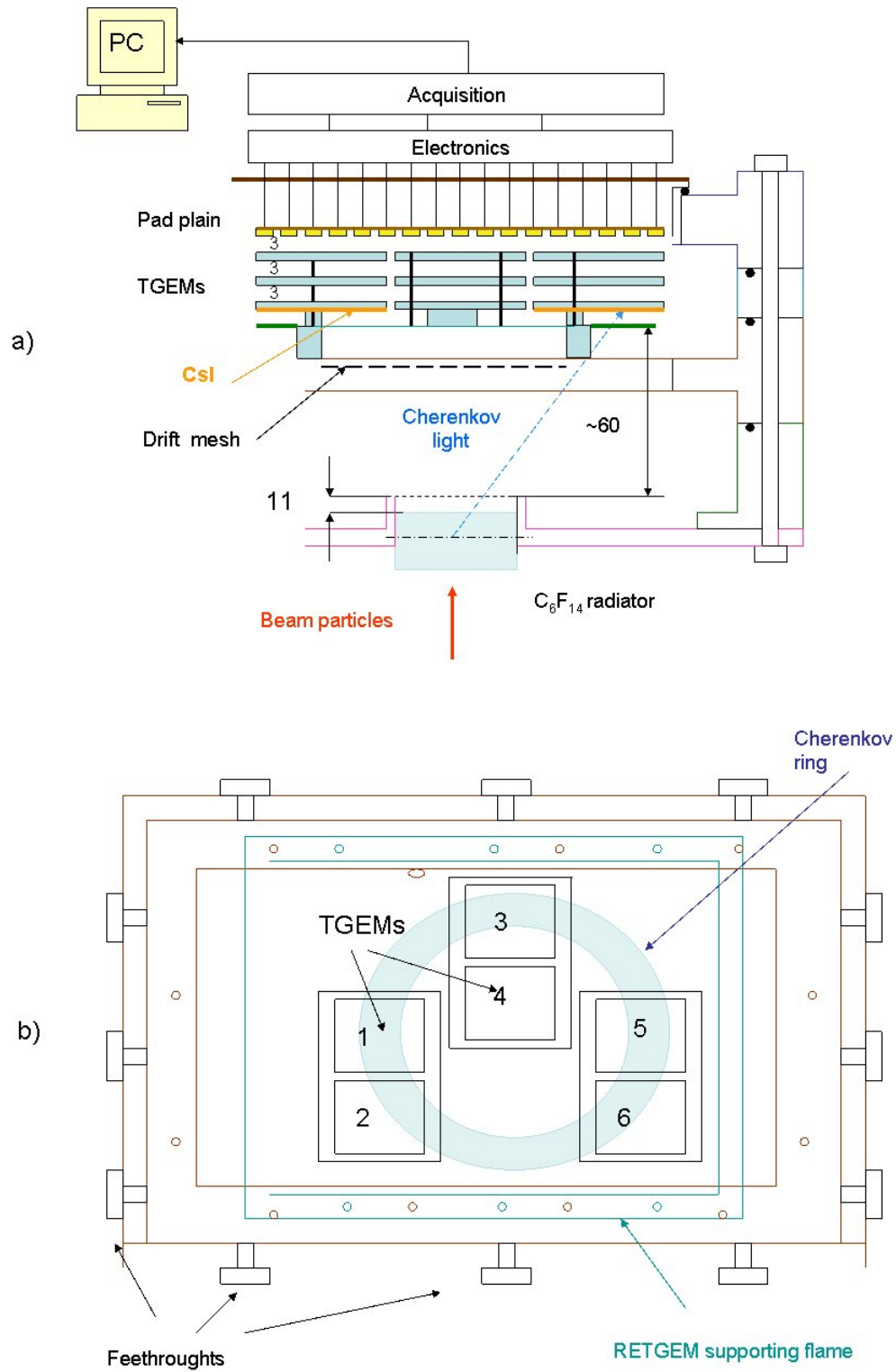

Figure 1. A schematic drawing of the RICH prototype: a) cross section, b) top view.

**Laboratory tests**

Before installation in the RICH detector, each TGEM was individually tested in a separate small gas chamber. In these tests we mainly identified the maximum achievable gains when the detectors were irradiated with a $^{55}$Fe source and with UV light. From 18 selected detectors six triple TGEMs were assembled on a supporting frame and transferred to the CERN CsI evaporation facility [9]. After coating them with a CsI layer, the quantum efficiency (QE) of the TGEM surfaces was measured in vacuum [9] and it was found to be slightly (~16%) below the QE of the MWPC currently used in the ALICE HMPID (due to the TGEM area covered by holes).

Afterwards, these six triple TGEMs were assembled using a glove box inside the RICH prototype's gas chamber.

We performed several laboratory tests of this prototype using UV light, 6 keV photons from $^{55}$Fe and electrons from $^{90}$Sr, such as: gas optimization, gas gain measurements in various conditions, identification of the sparkless zone of the TGEM, operation while detecting single photoelectrons and MIPs simultaneously, operational stability with time, etc. Most of the measurements were performed in Ne+10%CH$_4$ and Ne+10%CF$_4$ gas mixtures. As an example, the typical gas gain curves measured in these gases when the detector was simultaneously irradiated by UV and $^{55}$Fe are shown in Fig 2. As it can be seen, a gas gain of more than $10^5$ can be achieved in these conditions. At gains three times less than the achieved maximum no sparks were observed during 24 hours of continuous operation (with a rate per cm$^2$ of: 5x10$^4$ for UV and 4x10$^2$ for $^{55}$Fe). Several authors (see [4] and references therein) have observed instability of the TGEM gain in time: the gas gain can increase by a factor two during the first three to four hours of operation. We solved this problem by constantly keeping a certain voltage across the TGEMs (minimum 300 V). Under these conditions the gain variations were inside the few % interval. In another long-term stability test we operated the detector in Ne+10%CF$_4$ at a gas gain of $3\times10^5$ when it was simultaneously irradiated by UV and electrons from $^{90}$Sr. Not a single spark was observed for 24 hours of continuous operation (with a rate per cm$^2$ of: $5\times10^4$ for UV and $3.5\times10^4$ for $^{90}$Sr).

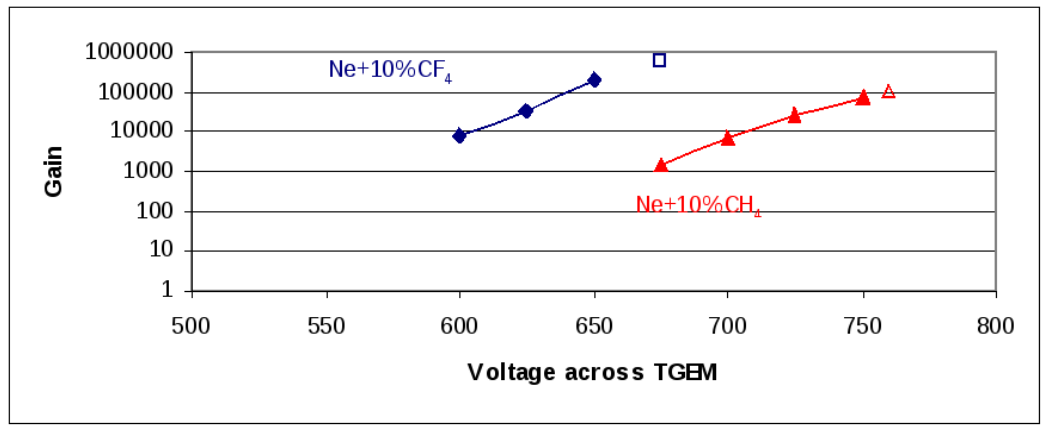

Figure 2. Results of gas gain measurements of triple TGEMs operated in Ne+10%CF$_4$ and Ne+10%CH$_4$ when the detector was simultaneously irradiated with UV light and 6 keV photons from $^{55}$Fe. Filled symbols represent measurements performed with a drift field of 170 V/cm and empty symbols correspond to measurements with a reversed drift field of 100 V/cm.

**Preliminary results from recent beam tests**

After the laboratory tests, the RICH prototype was installed in the T10 test beam facility at CERN. The trigger was provided by placing two pairs of finger scintillators, one pair in front and the other behind the chamber, defining a beam area of 1 cm$^2$. The ALICE DAQ framework [10] has been used for event recording, zero suppressed pad signals were read and stored for offline analysis. Figures 3 and 4 show the Cherenkov ring images (or more precisely, the charge signal from the pads) obtained in Ne+10%CH$_4$ and Ne+10%CF$_4$ gas mixtures, respectively. Besides the images of the Cherenkov ring we measured other various characteristics, for example: pulse height spectra, gas gain, cluster size (clusters are defined as pad patterns having all pads adjacent to another by at least one side), number of clusters per event and cetera.

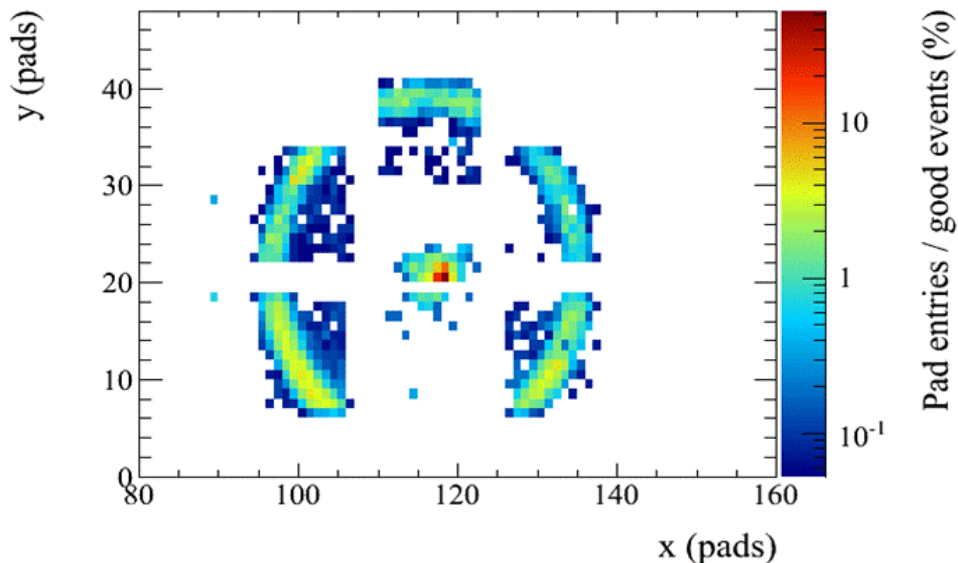

Figure 3. Detected Cherenkov ring (integrated events) in the recent November 2010 beam test at CERN in Ne+10%CH$_4$. The "spot" in the center represents MIPs (6 GeV/c pions) detected by the central TGEM.

Figure 5 shows the number of clusters per event measured with our two best TGEM stacks operating simultaneously. As can be seen, the mean number of clusters per event is around 1.82. In order to estimate the expected amount of clusters to be obtained in a large-area TGEM detector we should consider that the acceptance of these two TGEMs is ¼ of the total Cherenkov ring, so, in the case of total surface coverage we could expect to measure ~7 clusters per event. Also, the TGEM detection inhomogeneity has been extracted from experimental data, which accounts for variations in the CsI QE and the detection inefficiency at the TGEM borders (this effect can be easily reduced by increasing the size of the detectors). An inhomogeneity correction factor of ~0.78 has been calculated for the two TGEMs now under discussion. This gives us a total of ~9 raw clusters per event.

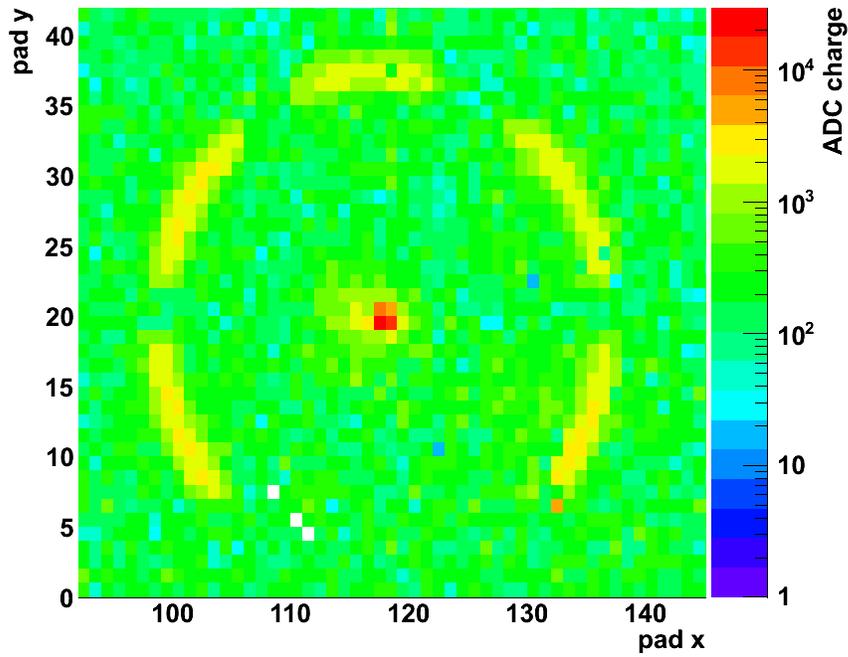

Figure 4. Cherenkov ring recorded in Ne+10%CF$_4$ (May 2011 beam test, raw data, no noise subtraction). The color scale represents pad charge in ADC channels (where 1 ADC channel ≃ 950 e$^-$).

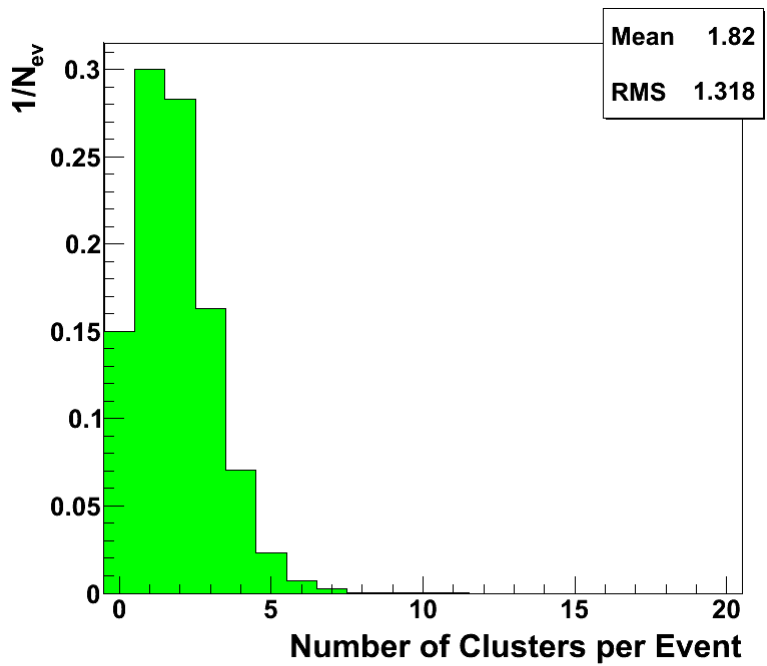

Figure 5. Number of clusters per event distribution measured with the two TGEM stacks that showed the best performance.

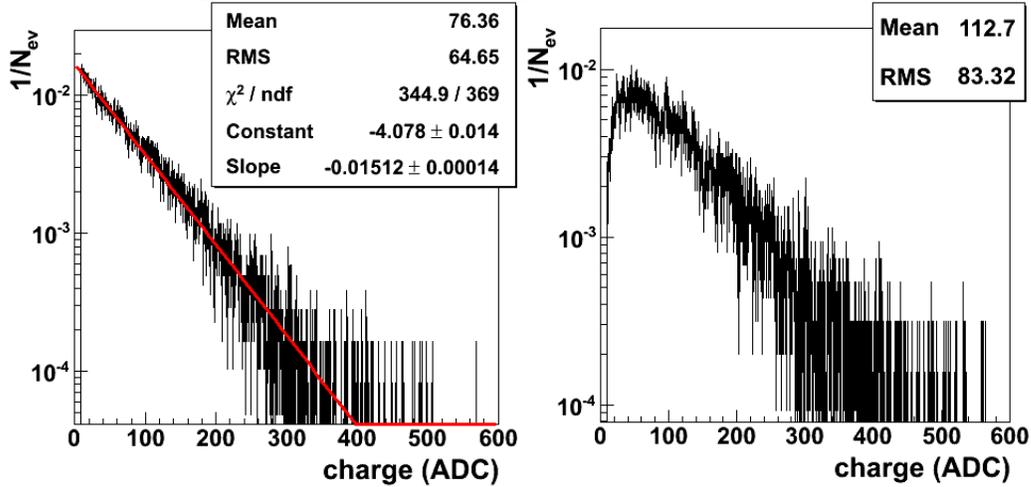

Figure 6. Charge distribution of clusters. Left: Distribution for clusters of size 1 (pad). An exponential fit has been done corresponding to a single photon spectrum. Right: Distribution for clusters of two pads or more. The behavior is not that of an exponential curve.

Testbeam data and Monte Carlo simulations have demonstrated that in the MWPC-based photon detector of the HMPID RICH counter [11] multiple photoelectrons are detected in the case of clusters larger than 2 pads. We expect a similar effect with the present prototype since two-pad clusters have double the average pulse-height with respect to single-pad clusters, which have the exponential distribution typical of single-photoelectron detection (Fig. 6). However, we must consider that with TGEMs the cluster topology is different with respect to the MWPC, for in the first case electrons are detected directly, while in the latter the measured signals are induced by ions moving towards the cathode, which result in larger clusters than what we have seen with the TGEM. Further work on this will give us the number of *resolved clusters* with which we can deduce the effective amount of photons detected. We should finally mention that the front-end electronics implemented in our test is not optimized for this application and could account for some inefficiency.

Note that also a computer program capable of performing the full scale simulations of our RICH prototype was developed by our group: starting from the conversion of Cherenkov photons to primary electrons, then the corresponding multiplication in the triple TGEMs and finally the signal formation delivered by the readout pad plane. As calculated by this program, the QE of the CsI TGEM photocathode is about 0.6 of the expected ideal conditions. Taking into account the complexity of the calculations and many approximations used, we consider the current results as very promising.

## Conclusions

We report the first successful implementation of a set of CsI-TGEMs with a liquid radiator where a Cherenkov ring has been observed. The results obtained, although modest, are encouraging and suggest that the present performance could be improved in the future by optimizing elements of the design.
The TGEM seems to be an interesting alternative to the MWPC technology especially since it allows the use of nonflammable gases.

**Acknowledgements;**
The authors would like to thank J. Van Beelen, M. Van Stenis and M. Weber for their help throughout this work.
G. Paic and D. Mayani acknowledge the support of the UNAM grant IN115808, the CERN-UNAM grant, the PAEP grant and the support of the Coordinacion de la Investigacion Cientifica and the Red de Fisica de Altas Energias.
G. Bencze was supported by Hungarian OTKA and NKTH grants NK77816, CK77719 and CK77815.